\documentclass[12pt,preprint]{article}%
\usepackage{amssymb}
\usepackage{amsmath}
\usepackage{graphics}
\usepackage{epsfig}
\usepackage{amsfonts}
\usepackage{graphicx}%
\renewcommand{\vec}[1]{{\bf #1}}
\newcommand{\eqb}{\begin{equation}}
\newcommand{\eqe}{\end{equation}}
\newcommand{\dmb}{\begin{displaymath}}
\newcommand{\dme}{\end{displaymath}}
\newcommand{\pd}{\partial}

\newcommand{\eab}{\begin{eqnarray}}
\newcommand{\eae}{\end{eqnarray}}

\newcommand{\e}{\mbox{e}}
\newcommand{\be}{\begin{equation}}
\newcommand{\ee}{\end{equation}}

\begin{document}
\begin{titlepage}
\begin{flushright}
\end{flushright}
\vspace{0.6cm}
\begin{center}
\Large{The quantum of action and finiteness of radiative corrections: Deconfining SU(2) Yang-Mills thermodynamics}
\vspace{1.5cm}\\ 
\large{Ralf Hofmann$^*$ and Dariush Kaviani$^\dagger$}
\end{center}
\vspace{1.0cm}
\begin{center}
{\em $^*$ Institut f\"ur Theoretische Physik\\
Universit\"at Heidelberg\\
Philosophenweg 16\\
69120 Heidelberg, Germany}
\end{center}
\vspace{0.5cm}
\begin{center}
{\em $^\dagger$ Institute for Particle Physics Phenomenology\\
Durham University\\
South Road\\
Durham, DH1 3LE, United Kingdom}
\end{center}
\vspace{1.0cm}
\begin{abstract}
The quantum of action $\hbar$, multiplying in certain powers perturbative vertices in 4D gauge 
theory, is related to the action of just-not-resolved selfdual and thermal gauge field configurations, calorons and anticalorons, 
of charge modulus unity. Appealing to the derivation of the effective theory for the deconfining 
phase of SU(2) Yang-Mills thermodynamics, we conclude that these vertex inducers 
convey a rapidly decreasing interaction strength between {\sl fundamental} plane waves when the 
momentum transfer is increased away from the scale of maximal, {\sl effective} resolution. This adds a deeper 
justification to the renormalization programme of perturbation theory which ignores the contribution to the partition function of nontrivially 
selfdual configurations. We also point out a connection between the QED fine-structure 
constant $\alpha$ and the electric-magnetically dual of the effective gauge coupling in the deconfining phase, 
and we illustrate the workings of effective loops in the expansion of the pressure.   
\end{abstract}
\end{titlepage}

\section{Introduction\label{intro}}

Key experiments in all areas of subatomic and condensed matter physics, 
spanning over more than a hundred years and designed to investigate aspects of the microscopic interaction between 
electromagnetic radiation, charged particles, and collective excitations, prove that the quantum of action $h=6.63\times 10^{-34}$\,Js, introduced by Max 
Planck in understanding the spectral radiance of a black-body cavity through eventual 
appeal to the statistical thermodynamics of Ludwig Boltzmann \cite{Planck1900}, is a {\sl universal} 
constant of Nature. Since its inauguration in 1900 the somewhat mysterious sneak-in of $h$, 
somehow being associated with `elementary regions' or `free rooms for action' 
\cite{PlancksNobelLecture}, into the 
conception and development of quantum mechanics \cite{Planck1900,DeBroglie,Einstein,Bohr,Heisenberg,Schrödinger} 
has been broaching a quest for deeper interpretation. 

To understand the meaning of $h$ as a mediator of indeterminism in fundamental interactions among 
propagating gauge fields, between gauge and charged matter fields, or in stationary binding physics could be of relevance 
in addressing as of yet poorly understood phenomena, for example those of 
2D strongly correlated electron systems \cite{HighTc}. Quantum Mechanics and perturbatively 
accessed Quantum Field Theory are highly successful, accurate, and efficient. However, both suffer 
from problems of mere interpretation. While the measurement process in Quantum Mechanics, that is, the interaction of 
a macroscopic observer with the subatomic system under investigation, mysteriously and in a 
nonlocal, instantaneous way collapses 
the evolved wave function of the latter to an eigenstate of the operator-valued, measured quantity 
the need for renormalization in Quantum Field Theory, in addition 
to not easing the classical Euler-Langrangian ambiguity inherent to the value of vacuum energy, is perceived by 
many as a necessary evil in putting perturbation theory to work. 
Fortunately, the predictive power of 4D gauge theories is not spoiled 
by the process of renormalization \cite{Veltman'Hooft,LeeZinnJustin}, and 
radiatively introduced gauge anomalies can be canceled granting the 
consistency of the highly successful Standard Model of Particle Physics. Yet one may wonder 
whether the subtraction of infinities in the perturbative 
loop diagrams of gauge theories enjoys a deeper justification possibly supplied 
by full nonabelian gauge dynamics.   

In the present paper we consider the (Euclidean) action 
of a fundamental, selfdual\footnote{Here the term `selfdual' stands for both negative and positive topological charge.} 
gauge field configuration of topological charge modulus unity (trivial-holonomy Harrington-Shepard \cite{HS1977} 
or nontrivial-holonomy Lee-Lu-Kraan-Van-Baal caloron/anti\-caloron \cite{LeeLuKraanVanBaal})
and scale parameter $\rho$ 
in the deconfining phase of thermalized SU(2) Yang-Mills theory. 
The action of a caloron/anti\-caloron, that is {\sl just not resolved} in the effective theory, 
is determined by the value of an effective coupling calculable 
in the according effective theory \cite{HerbstHofmann2004,Hofmann2005,Hofmann2007}. Recall 
that calorons and anticalorons are not resolved in this effective theory: 
Their contribution to effective dynamics fundamentally separates into a free 
part causing the emergence of an effective, inert, and spatially homogeneous 
adjoint scalar field $\phi$ \cite{HerbstHofmann2004} and the effects of interactions 
with the propagating plane-wave sector (vanishing topological charge). 
It is a fact that the energy-momentum tensor vanishes on any selfdual gauge-field configuration, 
see for example \cite{Hofmann2011}, and hence also on calorons and anticalorons. Since the latter configurations thus 
do not propagate one may wonder \cite{HS1977_II}
what their (potentially important) contribution to the thermodynamics of the 
high-temperature Yang-Mills gas actually is.   

For the generic nonabelian gauge group SU(2) 
this question was addressed in \cite{HerbstHofmann2004,Hofmann2005,Hofmann2007}, see also 
\cite{Hofmann2011} for a summarizing presentation: While the emergence of 
$\phi$ is sharply dominated by (large) calorons/anticalorons of radius $\rho$ smaller than but 
comparable\footnote{The integral over $\rho$ in the space-averaged two-point function of 
the fundamental field strength defining in singular gauge the kernel ${\cal K}$ of a differential operator $D$, with $\phi\in {\cal K}$, 
exhibits a cubic dependence on the cutoff when the latter is comparable 
to $|\phi|^{-1}$, see \cite{HerbstHofmann2004} and Sec.\,\ref{phiandrho} below.} 
to the scale of maximal resolution $|\phi|$ in the effective theory \cite{HerbstHofmann2004} interactions between calorons/anti\-calorons and 
plane waves manifest in a two-fold way effectively. First, there are small 
holonomy changes, induced through momentum transfers 
by plane-wave fluctuations that are largely off their 
free mass shell\footnote{All unresolved plane-wave fluctuations are at least by $|\phi|$ 
off their free mass shell.}, 
which generate negative ground-state pressure by virtue of 
a pure-gauge configuration of the effective gauge field \cite{Hofmann2005,Hofmann2011,Hofmann2011_I}. 
Second, as we shall make explicit in this paper, just-not-resolved 
calorons/anti\-calorons are responsible for pointlike 
vertices, associated with their dissociation into screened but unresolved 
monopole-antimonopole pairs on one hand \cite{SpatialWilson} and a rapidly converging loop expansion 
in the effective theory \cite{Schwarz2011,Kaviani2011,Hofmann2006,KavianiHofmann2007}. We will make clear that 
this loop expansion, which can naively be organized into an expansion in powers of $\hbar\equiv\frac{h}{2\pi}$, turns out to be a more 
subtle expansion. Still, the action of a just-not-resolved caloron or anticaloron, 
which introduces a vertex for effective plane waves, turns out to be $\hbar$.  

Calculating loops in effective variables yields a quantitative description of 
collective effects as induced by fundamental field configurations 
\cite{SpatialWilson,Resum2010,FalquezHofmann2011,LudescherHofmann2009}. 
Moreover, since the fields of the effective theory and their interactions do only depend on the contributions of 
large calorons/anticalorons we would conclude that, on a fundamental level, the mediation of interactions between unresolved 
off-shell plane-wave fluctuations by calorons/anticalorons (vertices) dies off rapidly with decreasing 
$\rho$. As a consequence, there are {\sl no} interactions with potentially large 
momentum transfers deep within the unresolved physics. We tend to interpret this as a justification for the perturbative 
renormalization procedure in nonabelian gauge theories. Finally, it is instructive to illustrate the technicalities of the effective loop 
expansion to contrast it with renormalized perturbation theory. 

The paper is organized as follows. Because our subsequent discussion relies on it 
we review in Sec.\,\ref{phiandrho} the emergence of an 
inert, adjoint scalar field $\phi$ by a spatial coarse-graining over noninteracting calorons and anticalorons of 
charge modulus unity within the deconfining phase of SU(2) Yang-Mills 
thermodynamics. For the same reason, we re-derive in Sec.\,\ref{tsc} the thermodynamically consistent 
solution for the temperature dependence of the effective gauge coupling $e$. In Sec.\,\ref{CP} we 
identify the Euclidean action of just-not-resolved calorons and anticalorons with the quantum of 
action $\hbar$ and deduce, by virtue of the dimensionlessness of the fine-structure 
constant $\alpha$, an interpretation of the fundamental unit of 
electric charge in terms of the dual to the effective SU(2) 
gauge coupling in the deconfining phase. Sec.\,\ref{UPW} discusses the implications of the action of a just-not-resolved 
caloron or anticaloron being associated with a local vertex. Namely, from the derivation of the effective 
theory we learn that such vertices are sharply suppressed with increasing momentum transfer 
(or decreasing caloron/anti\-caloron radius $\rho$) yielding, at {\sl any} externally prescribed resolution, 
finite answers for all radiative processes. Focusing on the example of QED, Sec.\,\ref{QED} 
discusses how just-not-resolved calorons and anticalorons of 
SU(2)$_{\tiny\mbox{CMB}}$ may induce local vertices between the plane waves belonging to electrically 
charged particles and propagating electromagnetic waves. To point out the contrast to 
renormalized perturbation theory, where the emergence of a scalar field associated with the 
maximal resolution and induced by the topologically nontrivial sector can not occur due to an a priori 
restriction to the plane-wave sector,  
Sec.\,\ref{Dariush} discusses the technicalities of the effective loop expansion of the 
pressure. Namely, we review the Feynman rules and constraints on loop variables and 
apply them to the (numerical) computation of three- and two-loop contributions 
to the pressure.

\section{Selfduality, propagating fields, and powers of $\hbar$\label{S1}}

\subsection{Inert adjoint scalar field $\phi$ and effective action\label{phiandrho}}

In \cite{HerbstHofmann2004} the following definition for the kernel ${\cal K}$ of the differential operator $D$ associated with $\phi$'s 
equation of motion in the deconfining phase of SU(2) Yang-Mills thermodynamics was shown 
to be unique:
\begin{equation}
\label{definition}
{\cal K}\equiv \{\hat{\phi}^a\}\equiv \sum_{C,A}\mbox{tr}\,\int d^3x \int d\rho
  \,t^a \, F_{\mu\nu} (\tau,\vec 0) \, \left\{ (\tau,\vec 0),(\tau,{\vec x})
\right\}\, F_{\mu\nu} (\tau,{\vec x}) \, \left\{ (\tau,{\vec x}),(\tau,{\vec 0})
\right\} \,,
\end{equation}
where
\begin{equation}  			
\label{abk}
\begin{split}
\left\{(\tau,\vec 0),(\tau,\vec x)\right\} 
&\equiv 
{\cal P} \exp \left[ i \int_{(\tau,\vec 0)}^{(\tau,\vec x)} dz_{\mu} \, A_{\mu}(z) \right]\,,  \\
\left\{(\tau,\vec x),(\tau,\vec 0)\right\} 
&\equiv 
\left\{(\tau,\vec 0),(\tau,\vec x)\right\}^\dagger\,,
\end{split}
\end{equation}
and $F_{\mu\nu}$ denotes the fundamental field-strength tensor. The Wilson lines in Eq.\,(\ref{abk}) are calculated along 
the straight spatial line connecting the points $(\tau,\vec 0)$ and $(\tau,\vec x)$, and 
${\cal P}$ demands path-ordering symbol. (The euclidean time coordinate is $\tau$.) 
In (\ref{definition}) the sum is over a 
noninteracting caloron ($C$) and a noninteracting anticaloron ($A$) of trivial holonomy and topological charge modulus unity, and 
$\rho$ denotes their (instanton) scale parameter. Both caloron 
and anticaloron are spatially centered at $\vec{z}_{C,A}=0$, for a discussion of the distribution of action density in 
dependence of $\rho/\beta$ ($\beta\equiv 1/T$, $T$ temperature) 
see \cite{GrossPisarskiYaffe}. In evaluating the integrals on the right-hand side of Eq.\,(\ref{definition}), one notices 
that only the magnetic-magnetic correlation gives rise to a 
nonvanishing result. Going over to spherical coordinates, both the unconstrained integration over $r\equiv|\vec{x}|$ and 
$\rho$ generate diverging multiplicative constants in front of a rapidly saturating sinusoidal time dependence separately 
arising from both the caloron 
as well as the anticaloron contribution. On the other hand, the azimuthal angular integration yields a 
muliplicative zero \cite{HerbstHofmann2004,Hofmann2011}. Together this generates an undetermined real prefactor 
which is one of the parameters spanning the kernel ${\cal K}$ of a uniquely 
determined second-order linear operator $D$. More specifically, one arrives at the following expression for the 
kernel ${\cal K}=\{\hat{\phi}^a\}$ of differential operator $D$
\eab
\label{calantical}
\{\hat{\phi}^a\}&=&\{\Xi_C(\delta^{a1}\cos\alpha_C+\delta^{a2}\sin\alpha_C)\,{\cal
  A}\left(2\pi(\hat{\tau}+\hat{\tau}_C)\right)\nonumber\\ 
&&+\Xi_A(\delta^{a1}\cos\alpha_A+\delta^{a2}\sin\alpha_A)\,{\cal
  A}\left(2\pi(\hat{\tau}+\hat{\tau}_A)\right)\}\ \ \ \ \ \ \ (a=1,2,3)\,,
\eae
where $\Xi_C$, $\Xi_A$, $\hat{\tau}_C$, and $\hat{\tau}_A$ are undetermined real parameters, and 
$\alpha_C$, $\alpha_A$ are angles that reflect the freedom of a global gauge choice. The amplitude function ${\cal A}$ 
is given as
\begin{equation}
\label{calA}
{\cal A}(2\pi\hat{\tau})=\frac{32\pi^7}{3}\int_0^{\xi} d\hat{\rho}\,\hat{\rho}^4\,
\left[\lim_{\hat{r}\to\infty}\sin(2\hat{g}(\hat{\tau},\hat{r},\hat{\rho}))\right]\frac{\pi^2\hat{\rho}^2+
\cos(2\pi\hat{\tau})+2}{\left(2\pi^2\hat{\rho}^2-\cos(2\pi\hat{\tau})+1\right)^2}\,,
\end{equation}
where $\xi$ represents a cutoff for the $\rho$ integration in units of $\beta$. (All quantities of 
dimension length are made dimensionless by division with $\beta$ and are denoted by a hat symbol). 
Function $\hat{g}$ saturates rapidly to a finite limit as $\hat{\rho},\hat{r}\to\infty$, the
integral over $\hat{\rho}$ in Eq.\,(\ref{calA}) diverges cubically for
$\xi\to\infty$, and function ${\cal A}(2\pi\hat{\tau})$ rapidly
approaches $\mbox{const}_\infty\times\xi^3\sin(2\pi\hat{\tau})$. 
Numerically, one obtains $\mbox{const}_\infty=272.018$. To summarize, the kernel ${\cal K}$, which uniquely 
determines operator $D$ as $D\equiv\partial^2_\tau+\left(\frac{2\pi}{\beta}\right)^2$, is sharply 
dominated by contributions of the $\hat{\rho}$ integration that lie in the vicinity of 
the upper limit $\xi$. 
 
Operator $D$ is explicitly 
dependent on temperature and thus, by itself, cannot be taken to be associated with a 
temperature-independent effective action\footnote{This action is not explicitly temperature dependent because the 
weights of selfdual configurations in the fundamental partition function are solely associated with 
their topological charges.} for the field $\phi$. The latter, however, arises by 
allowing a potential $V$ to absorb the explicit temperature dependence introduced by $D$. Appealing to the BPS 
condition inherited from the selfduality of the fundamental gauge-field configurations entering into the 
definitition (\ref{definition}), a first-order equation for $\phi$ is derived whose solution is parametrized 
by one constant of integration -- the Yang-Mills scale $\Lambda$. Perturbative renormalizability \cite{Veltman'Hooft,LeeZinnJustin}, constraining 
the interactions between effective plane-wave fields $a_\mu=a_\mu^a\,t_a$ ($t^a$ with $a=1,2,3$ the generators of SU(2) in fundamental representation, 
normalized to tr\,$t^a t^b=\frac12\delta^{ab}$), gauge invariance, and the fact that the field $\phi$ cannot, by definition, 
absorb or emit any energy-momentum yield the following unique result for the effective Euclidean action 
density ${\cal L}_{\mbox{\tiny eff}}$ \cite{HerbstHofmann2004,Hofmann2011}
\begin{equation}
\label{fullactden}
{\cal L}_{\mbox{\tiny eff}}[a_\mu]=\mbox{tr}\,\left(\frac12\,
  G_{\mu\nu}G_{\mu\nu}+(D_\mu\phi)^2+\frac{\Lambda^6}{\phi^2}\right)\,,
\end{equation} 
where $G_{\mu\nu}=\partial_\mu a_\nu-\partial_\nu
a_\mu-ie[a_\mu,a_\nu]\equiv G^a_{\mu\nu}\,t_a$ denotes the field 
strength for gauge field $a_\mu$, $D_\mu\phi=\partial_\mu\phi-ie[a_\mu,\phi]$, and $e$ is the effective
gauge coupling to be determined, see Sec.\,\ref{tsc}. Action 
density ${\cal L}_{\mbox{\tiny eff}}$ yields a highly accurate tree-level 
ground-state estimate and, as easily deduced in unitary gauge $\phi=2|\phi|\,t_3$, tree-level 
mass $m=2\,e|\phi|=2\,\sqrt{\frac{\Lambda^3}{2\pi T}}$ for propagating 
gauge modes $a_\mu^{1,2}$ rendering them thermal quasiparticles. 
   
\subsection{Thermodynamical consistency\label{tsc}}

For action density ${\cal L}_{\mbox{\tiny eff}}$ to be thermodynamically 
consistent implied quantities like pressure and energy density must be related by 
Legendre transformations. On the level of free thermal 
fluctuations\footnote{In the physical unitary-Coulomb gauge $\phi=2|\phi|\,t_3\,,\ \pd_i a_i^3=0$ the 
off-shellness of quantum fluctuations in the field $a_\mu$ is constrained by the maximal resolution 
$|\phi|$. Their contribution turns out to be negligible compared to 
that of thermal fluctuations \cite{Hofmann2005,Hofmann2011}.} we obtain the following evolution equation \cite{Hofmann2005}
 \begin{equation}
\label{evoleqsu2}
\partial_a\lambda=-\frac{24\lambda^4
  a}{(2\pi)^6}\frac{D(2a)}{1+\frac{24\lambda^3a^2}{(2\pi)^6}D(2a)}\,.
\end{equation}
Eq.\,(\ref{evoleqsu2}) is equivalent to \cite{GiacosaHofmann2007}
\begin{equation}
\label{evalambdasu2}
1=-\frac{24\lambda^3}{(2\pi)^6}\left(\lambda\frac{da}{d\lambda}+a\right)a\,D(2a)\,, 
\end{equation}
where $\lambda\equiv\frac{2\pi T}{\Lambda}$, $a\equiv\frac{m}{2T}$ and thus $a=2\pi e\lambda^{-3/2}$, 
and $D(y)\equiv \int_0^\infty dx\,\frac{x^2}{\sqrt{x^2+y^2}}\,\frac{1}{\e^{\sqrt{x^2+y^2}}-1}$. The right-hand side of 
Eq.\,(\ref{evoleqsu2}) is negative definite. As a consequence, its solution $\lambda(a)$ is strictly 
monotonic decreasing and so is its inverse $a(\lambda)$. Hence there is a regime in temperature $\lambda>\lambda_1$ 
where $a(\lambda)<1$, and Eq.\,(\ref{evalambdasu2}) can be approximated as 
\begin{equation}
\label{evalambdasu2sim}
1=-\frac{\lambda^3}{(2\pi)^4}\left(\lambda\frac{da}{d\lambda}+a\right)a\,. 
\end{equation} 
The solution to Eq.\,(\ref{evalambdasu2sim}), subject to the initial condition $a(\lambda_i)=a_i\ll 1$, is given as
\begin{equation}
\label{solasu2} 
a(\lambda)=4\sqrt{2}\pi^2\lambda^{-3/2}\left(1-\frac{\lambda}{\lambda_i}\left[1-\frac{a^2_i\lambda_i^3}{32\pi^4}\right]\right)^{1/2}\,.
\end{equation}
Thus for $\lambda_1<\lambda\ll\lambda_i$ function $a(\lambda)$ runs into the
attractor $a(\lambda)=4\sqrt{2}\pi^2\lambda^{-3/2}$. Using $a\equiv\frac{m}{2T}=2\pi
e\lambda^{-3/2}$, this attractor corresponds to the plateau $e\equiv\sqrt{8}\pi$. Because the
attractor increases with
decreasing $\lambda$ the condition $a\ll 1$ will be violated at small 
temperatures. The estimate $14.61>\lambda_1$ is obtained by setting the 
attractor equal to unity. Lowering $\lambda$ further, the solution to the 
full Eq.\,(\ref{evalambdasu2}) runs into a thin pole at $\lambda_c$ of the form \cite{Hofmann2011}
\eqb
\label{lambdapole}
e(\lambda)=-4.59\,\log(\lambda-\lambda_c)+18.42\,,
\eqe
where $\lambda_c=13.87$. Now, setting the upper integration limit 
$\xi$ in Eq.\,(\ref{calA}) equal to $1/(\beta |\phi|^{-1})$ and invoking 
$\lambda_c=13.87$, yields $\xi\ge (13.87)^{3/2}/(2\pi)\sim 8.22$. 
But the approach of function ${\cal A}(2\pi\hat{\tau})$ to a sine is 
perfect for $\xi\sim 8.22$ \cite{HerbstHofmann2004,Hofmann2011}, and 
for $\xi\sim 8.22$ the integral in Eq.\,(\ref{calA}) is 
strongly dominated by its upper limit. We conclude 
that calorons and anticalorons that are just not 
resolved dominate the emergence of the field $\phi$ and 
its consequences for the effective gauge dynamics 
(adjoint Higgs mechanism and thermal ground state estimate).     

The divergence of $e$ at $\lambda_c$ initiates a transition to a phase 
where thermal quasiparticles decouple, the ground state is a condensate of magnetic monopoles, and 
a formerly massless gauge mode acquires mass by the dual Meissner effect 
\cite{Hofmann2005,Hofmann2011}.  

We conclude that {\sl almost everywhere} 
in the deconfining phase the effective coupling assumes the constant value 
$e\equiv\sqrt{8}\pi$. 

\subsection{Counting of powers of $\hbar$ and action of just-not-resolved caloron\label{CP}}

Concerning the counting of powers in $\hbar$, 
loop expansions in the effective theory are carried out as in conventional 
perturbation theory. Additional subtleties arise because of the existence 
of a maximal resolution $|\phi|$ which enforces constraints on momentum 
transfers in four-vertices and on the off-shellness of the massless 
mode. These constraints have an obvious form in physical unitary-Coulomb gauge, 
see Sec.\,\ref{Dariush}. To make the power counting in $\hbar$ most explicit, we work in 
units where $k_B=c=\epsilon_0=\mu_0=1$ but $\hbar$ is re-instated as an action. (So far we have 
worked in supernatural units $\hbar=c=k_B=1$). 
The (dimensionless) exponential 
\eqb
\label{expweight}
-\frac{\int_0^{\beta} d\tau d^3x\,{\cal L}^\prime_{\mbox{\tiny eff}}[a_\mu]}{\hbar}\,, 
\eqe
in the weight belonging to fluctuating fields in the partition function\footnote{Recall that $\phi$ is inert. 
As a consequence, the factor, whose exponent is the potential-part of the effective action, can be pulled out of the partition 
function and needs not be considered in a discussion of the effective loop expansion.} thus can, in unitary gauge, be re-cast as 
\eqb
\label{expweightdiml}
-\int_0^{\beta} d\tau d^3x\,\mbox{tr}\,
\Big(\frac12(\partial_\mu \tilde{a}_\nu-\partial_\nu
\tilde{a}_\mu-ie\sqrt{\hbar}[\tilde{a}_\mu,\tilde{a}_\nu])^2-e^2\hbar[\tilde{a}_\mu,\tilde{\phi}]^2\Big)\,,
\eqe
where $\tilde{a}_\mu\equiv a_\mu/\sqrt{\hbar}$ and $\tilde{\phi}\equiv \phi/\sqrt{\hbar}$ are assumed to not depend on $\hbar$ \cite{Brodsky2011}  , 
see also \cite{DonoghuePapers}. Notice that because of the terms  
$\propto \hbar^0$ in (\ref{expweightdiml}) the unit of $\tilde{a}_\mu$ is 
length$^{-1}$. Thus the coupling $e$ must have the unit of $1/\sqrt{\hbar}$ 
(and the unit of $\tilde{\phi}$ is also length$^{-1}$). Together with the results reviewed 
in Sec.\,\ref{tsc} we thus have
\eqb
\label{plateaucoupl}
e=\frac{\sqrt{8}\pi}{\sqrt{\hbar}}
\eqe
almost everywhere in the deconfining phase. 
Because fundamental field configurations are smoothly connected 
to the effective dynamics\footnote{The spatial coarse-graining rapidly saturates: 
$\frac{|\phi|^{-1}}{\beta}\ge
8.221\times\left(\frac{\lambda}{\lambda_c}\right)^{3/2}$ and practically no alteration of $\phi$'s 
time dependence occurs when varying $r,\rho\sim |\phi|^{-1}$ \cite{Hofmann2011}.} the action $S_{C,A;\rho\sim |\phi|^{-1}}$ of a 
just-not-resolved caloron/anti\-caloron reads
\eqb
\label{actioncalantcal}
S_{C,A;\rho\sim |\phi|^{-1}}=\frac{8\pi^2}{e^2}=\hbar\,.
\eqe
Equation (\ref{actioncalantcal}) has implications. First, it suggests that the quantum of action 
$\hbar$, whose introduction in (\ref{expweight}) is motivated by the laws of Quantum Mechanics and which should enable a systematic accounting of quantum 
corrections (number of loops) in the effective theory, coincides with 
the Euclidean action of a just-not-resolved selfdual (classical) field configuration whose 
topological charge is fixed to be of modulus 
unity \cite{HerbstHofmann2004,Hofmann2011}. That is, the reason why 
$\hbar$ really is a constant can be traced to the universal constancy of $e$ (almost no dependence on $T$ and 
no dependence on the specific Yang-Mills scale $\Lambda$ of a given 
SU(2) gauge theory) and the 
fact that only charge-modulus-unity calorons/anticalorons contribute to the 
effective thermodynamical consistency of interacting, fundamental 
topological and plane-wave configurations \cite{HerbstHofmann2004}. Moreover, rather than multiplying first and second 
powers of $\hbar^{1/2}$ onto the powers $e$ and $e^2$ 
of an a priori unknown coupling constant $e$ in three- and four-vertices of a Yang-Mills 
theory we would infer that these vertices exist only because of the presence of just-not-resolved 
calorons and anticalorons. Second, for the fine-structure constant $\alpha$ of 
Quantum Electrodynamics (QED) to be dimensionless,  
\eqb
\label{FSC}
\alpha=N^{-1}\frac{g^2}{4\pi\hbar}\,,
\eqe
the coupling $g$ in Eq.\,(\ref{FSC}) must have the unit of $\sqrt{\hbar}$. ($N^{-1}$ denotes a numerical 
factor related to the mixing of the massless modes belonging to several SU(2) 
groups, see \cite{Hofmann2011,GiacosaHofmann2005}.) This is guaranteed if $g$ is taken to be the electric-magnetically dual coupling 
to $e$: 
\eqb
\label{elch}
g=\frac{4\pi}{e}\propto \sqrt{\hbar}\,.
\eqe
That is, a magnetic monopole liberated by the dissociation of a large-holonomy SU(2) caloron and other incarnations of this 
magnetic charge in the preconfining and confining phase\footnote{For example, the isolated charge situated in the center of the flux 
eddy associated with the selfintersection of a center-vortex loop in the confining phase \cite{Hofmann2011,MoosmannHofmann2008}, 
see also discussion in Sec.\,\ref{QED}.} are interpreted as {\sl electric} charges in the real world. Third, 
by virtue of Eq.\,(\ref{plateaucoupl}) the factors $e\sqrt{\hbar}$ and $e^2\hbar$ in (\ref{expweightdiml}) 
do not depend on $\hbar$. Thus, by assumption the weight of the partition function associated with the effective theory is independent of 
$\hbar$, and thus the effective loop expansion is {\sl not} an expansion in powers of $\hbar$. This is in contrast to renormalized perturbation 
theory where the value of the coupling is a priori unknown: The perturbative coupling constant needs to 
be determined by an empirical boundary condition. 
     
\subsection{Unresolved, interacting plane waves: Local vertices by caloron/anticaloron mediation\label{UPW}}

From the results of Sec.\,\ref{CP} and the facts reviewed in Sec.\,\ref{intro} we may conclude that in the effective 
theory the pure-gauge configuration, associated with the a priori estimate of the thermal ground state, 
describes the collective effect of interacting (typically small-holonomy) calorons/anticalorons of 
action $\hbar$ that are just not resolved. Recall that the holonomy of calorons/anticalorons is elevated from small to large by effective plane-wave 
interactions which collectively account for dissociation into screened but unresolved 
magnetic monopole-antimonopole pairs \cite{SpatialWilson,Diakonov2004}. Extrapolating the finding, 
that it is unresolved calorons/anticalorons which induce local vertices 
between plane waves, deeply into the unresolved domain, we would by the (ultraviolet as well as infrared) 
finiteness of all quantities calculable in the effective theory and the fact that the emergence of the field 
$\phi$ is dominated by just-not-resolved calorons/anticalorons reason (cubic dependence of the prefactor of $\phi$'s 
time dependence on the cutoff for the $\rho$ integration, see Eq.\,(\ref{calA}) and \cite{HerbstHofmann2004,Hofmann2011}) 
that the strength of interactions between 
fundamental plane waves, which are largely off their 
free mass shell, rapidly ceases with increasing momentum transfer. Recall that the effective pure-gauge configuration 
of the thermal ground-state estimate, 
which collectively describes these interactions, 
is {\sl sourced} by the field $\phi$ \cite{Hofmann2005,Hofmann2007,Hofmann2011}. 
But again, the content of fundamental charge-modulus unity selfdual gauge-field configuration in 
$\phi$ is {\sl strongly dominated} by those calorons/anti\-calorons that are {\sl just not resolved}.    

\begin{figure}
[ptb]
\begin{center}
\includegraphics[
height=2.124in,
width=5.6628in]{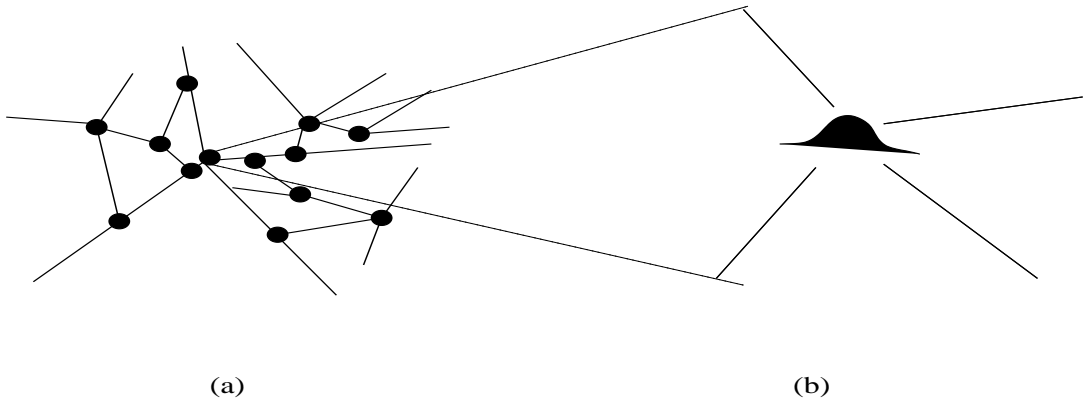}
\caption{(a) Plane waves (lines) of a Yang-Mills theory interacting via local vertices (solid circles), (b) generation of a vertex 
(local crumble within the ensemble of noninteracting plane waves) 
by a nonpropagating and selfdual gauge-field configuration of topological charge modulus unity in the same Yang-Mills theory.}
\label{Fig-1}
\end{center}
\end{figure}

The point of view, see Fig.\,\ref{Fig-1}, that the emergence of local vertices between plane waves in a Quantum Yang-Mills theory 
is a consequence of the existence of nonpropagating, topologically stabilized and selfdual Euclidean 
field configurations of the same theory, compares in an interesting way to pure perturbation theory 
where such local vertices are taken for granted \footnote{For the effective and local interaction between fundamentally 
charged chiral fermions via their zero modes, which are localized on instantons, the 't Hooft vertex represents 
an example of plane waves interacting by the mediation of selfdual gauge-field configurations \cite{'tHooft1976I,'tHooft1976II}.}. Namely, while perturbation theory requires a 
renormalization programme to sweep ultraviolet divergences under the carpet the latter 
do not occur to begin with when vertices are understood to be induced by unresolved, nonpropagating, topological, 
and selfdual field configurations: Because potential plane-wave fluctuations, that are extremely off-shell compared 
to an externally imposed resolution\footnote{In the case of thermodynamics this resolution is 
prescribed by temperature to be $|\phi|$.}, do not interact they are not generated in 
the first place. To the authors' minds this situation does by no means contradict the renormalization programme 
of perturbation theory but rather supplies it with a deeper justification: The subtraction of 
infinities in perturbative loop diagrams, which ignore the presence of 
the nontrivially selfdual sector of gauge-field configurations, simulates the nonoccurrence of 
such divergences in the full Yang-Mills theory.          
    
\subsection{Gauge-field-matter interactions\label{QED}}

So far we have discussed the situation of a pure SU(2) Yang-Mills theory. The undisputed 
successes of perturbatively accessed QED in particular (and the Standard Model of Particle Physics as a whole) demand to put the results of 
Secs.\,\ref{CP} and \ref{UPW} into the perspective of electrically charged particle interacting with photons. 
Let us undertake a perhaps speculative discussion of the situation in QED. 

To attribute caloron/anti\-caloron mediation to a QED vertex, the U(1) gauge group must be considered a dynamically broken 
SU(2). As explained in Sec.\,\ref{intro} and investigated in \cite{HerbstHofmann2004,Hofmann2005,Hofmann2007,Hofmann2011} 
this gauge-symmetry breaking is collectively mediated by calorons 
and anticalorons in terms of the field $\phi$. In \cite{Hofmann2009} the critical temperature 
$T_c$ of this SU(2) theory (and thus its Yang-Mills scale) was, by interpretation of the observed low-frequency 
excess in line temperature (see references in \cite{Hofmann2009}), 
extracted to be the present baseline temperature 2.725\,Kelvin 
of the Cosmic Microwave Background (CMB). Thus the name SU(2)$_{\tiny\mbox{CMB}}$. 

Radiative corrections in QED are organized 
into a power series in the fine-structure constant 
$\alpha$. As already mentioned in Sec.\,\ref{CP} $\alpha$ measures the strength of interaction 
of the massless mode of SU(2)$_{\tiny\mbox{CMB}}$ (the propagating photon) with a screened yet unresolved 
monopole \cite{SpatialWilson}. Such a monopole becomes isolated and 
thus resolvable when trapped in the core of the selfintersection 
of a center-vortex loop \cite{MoosmannHofmann2008II}. Notice that in the presence of propagating photons 
a center-vortex loop of selfintersection number unity 
is the only absolutely stable degree of freedom in the confining phase of an 
SU(2) Yang-Mills theory. (The nonselfintersecting species shrinks to a round point \cite{MoosmannHofmann2008}
and vortex loops of intersection number larger than unity decay by charge annihilation or 
repulsion \cite{MoosmannHofmann2008II,Hofmann2007CVL}.) For the SU(2)$_{\tiny\mbox{CMB}}$ photon to see 
the charge of a monopole belonging to another SU(2) theory, which presently is in its confining phase, 
the two theories must mix. So the crumble in the photonic plane wave, induced by the 
presence of the electric charge of, say, an electron, occurs by having the latter couple 
to its very massive photon (confining phase) which in turn mixes into the SU(2)$_{\tiny\mbox{CMB}}$ plane wave 
by caloron/anti\-caloron mediation of the latter theory. Recall that, 
according to the results of Sec.\,\ref{CP}, this mediation is accompanied 
by a unit $\hbar$ of `free rooms for action'.     

Interestingly, since the QED action, involving the electron field $\psi$ and $a_\mu^3$, is given as \cite{Brodsky2011}
\eqb
\label{QEDact}
S_{\tiny\mbox{QED}}=\int d^4x\,\left(\bar{\psi}(\gamma^\mu(i\hbar\partial_\mu-g\,a_\mu^3 )-m\right)\psi\,,
\eqe
where now $\psi$, $\tilde{m}\equiv\frac{m}{\hbar}$, and $\tilde{a}_\mu^3$ are considered independent 
of $\hbar$, one arrives at
\eqb
\label{QEDactscal}
-\frac{S_{\tiny\mbox{QED}}}{\hbar}=-\int d^4x\,\left(\bar{\psi}(\gamma^\mu(i\partial_\mu-\frac{1}{\sqrt{\hbar}} g\,a_\mu^3 )-\tilde{m}\right)\psi\,.
\eqe
By virtue of Eq.\,(\ref{elch}) we would thus conclude from Eq.\,(\ref{QEDactscal}) that the interaction $\frac{1}{\sqrt{\hbar}} g
\bar{\psi}\gamma^\mu a_\mu^3\psi$ does not exhibit an $\hbar$ dependence. So even in 
QED the loop expansion can no longer be considered an expansion in powers of $\hbar$ if the 
vertex is interpreted to be induced by a caloron/anticaloron of SU(2)$_{\tiny\mbox{CMB}}$.

\section{Effective and irreducible, planar 3-loop corrections\label{Dariush}}

After `scratching into unresolved physics' in Sec.\,\ref{S1} to associate interactions 
between plane waves with the presence of calorons/anti\-calorons we would here 
like to re-iterate \cite{KavianiHofmann2007,HerbstHofmannRohrer2004,SchwarzHofmannGiacosa2007} on technicalities in performing the effective loop expansion 
of the pressure in deconfining SU(2) Yang-Mills thermodynamics \cite{Hofmann2006}.

\subsection{Feynman rules and constraints}

Following the extraction of the effective gauge coupling $e(\lambda)$ 
from the thermodynamical consistency of effective, noninteracting quasiparticles, 
one may investigate the magnitude of radiative effects implied by the so-obtained function 
$e(\lambda)$. This strategy turns out to be selfconsistent. To perform the 
calculation we go back to supernatural units and work within the 
real-time formulation \cite{Hofmann2005,Hofmann2007,Hofmann2011}. 
Thus $\tilde{a}_\mu\to a_\mu$ etc.      

The Feynman rules are formulated in the completely fixed (and physical) unitary-Coulomb gauge which avoids the consideration of 
Faddeev-Popov ghosts. 
$\\$
$\bullet$ For the three-vertex $\Gamma^{\mu\nu\rho}_{[3]abc} (p,k,q)$ and for the 
four-vertex $\Gamma^{\mu\nu\rho\delta}_{[4]abcd}(p,q,r,s)$, we read off from Eq.\,(\ref{expweightdiml}) that
\begin{eqnarray}
\label{D1}
\Gamma^{\mu\nu\rho}_{[3]abc} (p,k,q) & = & e(2\pi)^4 \delta (p+q+k) \varepsilon_{abc}[g^{\mu\nu}(q-p)^{\rho}+g^{\nu\rho}(k-q)^{\mu}+ \notag\\[1ex]
&& g^{\rho\mu}(p-k)^{\nu}], \\ \Gamma^{\mu\nu\rho\delta}_{[4]abcd}(p,q,r,s) & = & -ie^2(2\pi)^4\delta(p+q+s+r)
[\varepsilon_{fab}\varepsilon_{fcd}(g^{\mu\rho}g^{\nu\sigma}-g^{\mu\sigma}g^{\nu\rho})+ \nonumber  \\
&& \varepsilon_{fac}\varepsilon_{fdb}(g^{\mu\sigma}g^{\rho\nu}-g^{\mu\nu}g^{\rho\sigma})+ 
\varepsilon_{fad}\varepsilon_{fbc}(g^{\mu\nu}g^{\sigma\rho}-g^{\mu\rho}g^{\sigma\nu})]\,,\nonumber\\ 
\end {eqnarray}
where $\varepsilon_{abc}$ are the structure constants of SU(2). 
$\\$ 
$\bullet$  The propagator of the tree-level heavy (TLH) modes propagates thermalized on-shell 
particles (no energy-momentum tranfer to $\phi$ in the summation of the 
Dyson series for the interaction of $a_\mu^{1,2}$ with $\phi$) of mass $m=2\,e|\phi|$ only \cite{Hofmann2011}
\begin{eqnarray}
\label{D2}
D^{\text{TLH}}_{\mu\nu,ab}(p) & = & -2\pi\delta_{ab}\tilde D_{\mu\nu}\delta(p^2-m^2)n_B\left(\frac{|p^0|}{T}\right)\,, \\
\tilde D_{\mu\nu} & = & \bigg(g_{\mu\nu}-\frac{p_{\mu}p_{\nu}}{m^2}\bigg)\,.
\end{eqnarray}
The propagation of tree-level massless (TLM) modes splits into a vacuum and a thermal part:
\begin{eqnarray}
\label{D3}
D^{\text{TLM}}_{\mu\nu,ab}(p) &=& -\delta_{ab}\left\{P^{T}_{\mu\nu}\left[\frac{i}{p^2} + 
2\pi\delta (p^2)n_B\left(\frac{|p^0|}{T}\right)\right] - \frac{iu_{\mu}u_{\nu}}{\vec{p}^2}\right\}\,,
\end{eqnarray}
where 
\begin{equation}
\label{D4}
P^{00}_T = P^{0i}_T = P^{i0}_T \equiv 0,\;\;\;\ P^{ij}_T = \delta^{ij} - \frac{p^{i}p^{j}}{\vec{p}^2}\,.
\end{equation}
Here $n_B(x)=\frac{1}{e^x-1}$ denotes the Bose-Einstein distribution function. Notice that TLM modes have a color index $a=3$ 
and TLH modes $a=1$ or $a=2$. Moreover, the term $\propto u_{\mu}u_{\nu}$ in Eq.\,(\ref{D3}) expresses the instantaneous 
``propagation'' of the $a^3_0$ field (Coulomb term\footnote{However, one-loop radiatively induced, the emergence of 
longitudinally propagating $E$-field (charge-density) waves is observed \cite{FalquezHofmann2011}.}), and $u_{\mu} = (1,0,0,0)$ represents the four-velocity of the heat bath.
$\\$
$\bullet$ The maximal off-shellness of a TLM mode with four-momentum $p$ is constrained as \cite{Hofmann2005,Hofmann2006,Hofmann2011}:
\begin{equation}
\label{D5}
|p^2|  \leq  |\phi|^2 \quad  \mbox{(TLH modes)}\quad\quad |p^2| \leq  |\phi|^2  \quad \mbox{(TLM modes)}\,.
\end{equation}
For a four-vertex with ingoing four-momenta $p_1,p_2$ and outgoing four-momenta $p_3,p_4$ 
we have \cite{Hofmann2007,Hofmann2006,Hofmann2011}
\begin {eqnarray}
\label{D8}
|(p_1 + p_2)^2| &\leq& |\phi|^2 \quad \mbox{(s-channel)}, \quad |(p_3 - p_1)^2| \leq |\phi|^2 \quad \mbox{(t-channel)}, \notag \\
|(p_3 - p_2)^2| &\leq& |\phi|^2 \quad \mbox{(u-channel)}.
\end{eqnarray}
As we shall review \cite{Hofmann2006,KavianiHofmann2007} in Sec.\,\ref{GELC}, 
conditions (\ref{D5}), (\ref{D8}), and the requirement of on-shellness of TLH modes imply a rapid growth of the 
number of constraints over the number of a priori noncompact (radial) integration variables with an increasing number of 
loops in an irreducible diagram.

\subsection{Expectations for convergence properties of loop expansion from 
two- and three-loop corrections to the pressure \label{GELC}}

\begin{figure}
[ptb]
\begin{center} 
\includegraphics[height=1.63in, width=4.5in]{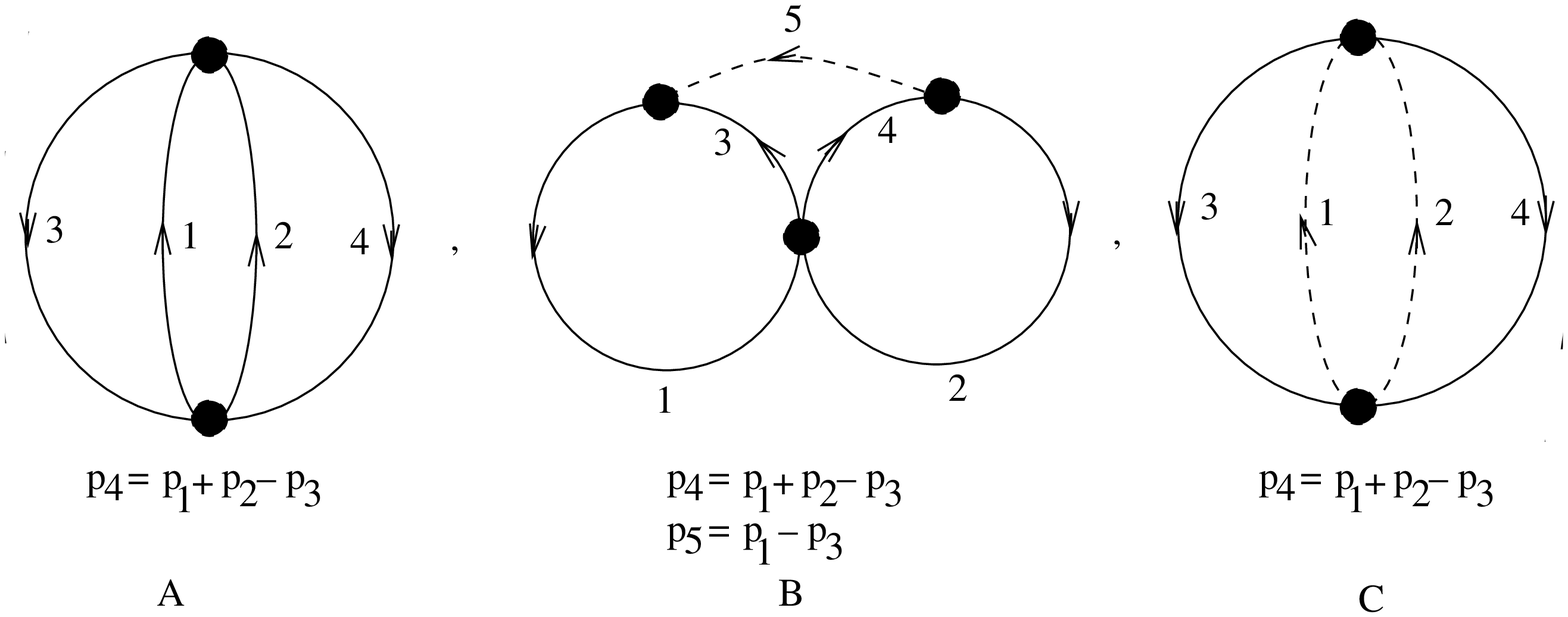}
\caption{The three irreducible, planar three-loop diagrams with TLH  (solid) and TLM (dashed) modes in the loops.}
\label{Fig-3}
\end{center}
\end{figure}

As discussed in the previous section, the momentum transfer in all Mandelstam variables s,t, and u 
in a four-vertex with ingoing four-momenta $p_1, p_2$ and outgoing 
four-momenta $p_3, p_4$ should not exceed the scale $|\phi|$. 
For three-loop diagrams, and in particular the ones of Fig.\,\ref{Fig-3}, where the cutting of any 
line yields a 1Pi irreducible contibution to a polarization tensor, 
the number $\tilde{K}$ of independent, radial loop-momentum variables 
$(p_{0},|\vec{p}|)_{i}$ for $i=1,2,3$ is $\tilde{K}=6$. To judge the 
number $K$ of independent constraints imposed on them we, 
in addition to (\ref{D5}) and (\ref{D8}), to consider on-shellness constraints for the TLH modes: 
\begin{equation}
\label{D10}
p_1^2=m^2,\;\;\;\ p_{2}^2=m^2,\;\;\;\ p_{3}^2=m^2,\;\;\;\,p_{4}^2=(p_{1}+p_{2}-p_{3})^2=m^2\,.
\end{equation}
Conditions (\ref{D8}) and (\ref{D10}) together yield for irreducible three-loop diagrams A and B a total of $K=3+4=7$ constraints. 

According to (\ref{D5}), the TLM modes in the irreducible three-loop diagram C are subject to the maximal off-shellness conditions
\begin {equation}
\label{D11}
|p_1^2| \leq |\phi|^2,\;\;\;\ |p_2^2| \leq |\phi|^2\,,
\end{equation}
and the on-shellness relations for the TLH modes in diagram C are  
\begin{equation}
\label{D12}
p_3^2=m^2,\;\;\;\,p_{4}^2=(p_{1}+p_{2}-p_{3})^2=m^2.
\end{equation}
Thus, again, we have $K=3+4=7$ constraints for diagram C. 
Hence for all irreducible three-loop diagrams we have $K=7$.
Therefore, the number of constraints exceeds the number of radial variables:
\begin{equation}
\label{D13}
\tilde{K}<K\,.
\end{equation}    
This relation implies that irreducible three-loop integrations have either 
compact or empty supports. 

\begin{figure}
[ptb]
\begin{center} 
\includegraphics[height=1.63in, width=3.5in]{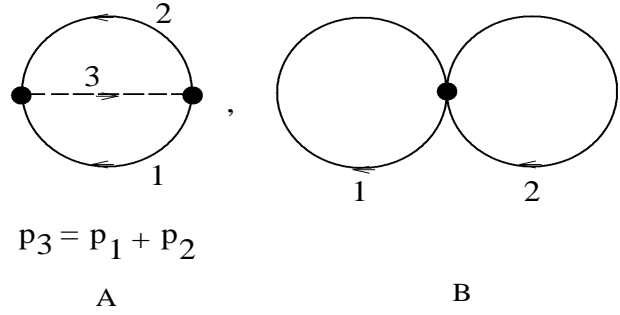}
\caption{Two out of three nonvanishing two-loop diagrams with TLH  (solid) and TLM (dashed) modes in the loops.}
\label{Fig-2}
\end{center}
\end{figure}

For the two-loop diagrams of Fig.\,\ref{Fig-2}, we have $\tilde{K}>K$. 
Namely, the number of $\tilde{K}$ of independent radial 
loop momenta $(p_{0},|\vec{p}|)_{i}$ for $i=1,2$ is $\tilde{K}=4$ whereas the constraints plus the on-shellness conditions are
\begin{equation}
\label{D14}
|(p_{1}+p_{2})^2| \leq |\phi|^2,\;\;\;\ p_1^2=m^2,\;\;\;\ p_2^2=m^2.
\end{equation}
This implies $K=1+2=3$. Thus, unlike the case of irreducible three-loop integrations, we expect the support of two-loop 
integrations to be noncompact. In Ref.\,\cite{Hofmann2006} a counting of $\tilde{K}$ versus 
$K$ was performed for an arbitrary loop number in the case of planar and nonplanar 
irreducible loop diagrams which gives rise to the 
conjecture that, modulo resummations of irreducible contributions to the 
polarization tensors in the lines of a given bubble diagram, 
the expansion of the pressure {\sl terminates} at a finite loop order. 
The fact that $\tilde{K}<K$ \cite{KavianiHofmann2007} is already in agreement with this conjecture. 

In the remaining subsections of Sec.\,\ref{Dariush} 
we discuss how numerical results are obtained and that, compared to the free quasiparticle pressure, two-loop and 
irreducible three-loop diagrams exhibit a hierarchical suppression of their contributions to the thermodynamical 
pressure \cite{KavianiHofmann2007}. In fact, diagram C of Fig.\,\ref{Fig-3} is shown to vanish. 

\subsection{Irreducible 3-loop integrations}

For the irreducible three-loop bubble diagrams in Fig.\,\ref{Fig-3}, we have in case A 
two four-vertices and four propagators and the correction to the pressure is
\begin{eqnarray}
\label{D15}
\Delta P_A &=& \frac{1}{48}\int \frac{d^4p_1d^4p_2d^4p_3}{(2\pi)^4(2\pi)^4)(2\pi)^4} \Gamma_{[4]abcd}^{\mu\nu \rho \sigma} 
\Gamma_{[4]\bar{a}\bar{b}\bar{c}\bar{d}}^{\bar{\mu}\bar{\nu} \bar{\rho}\bar{\sigma}}D_{\rho \bar{\rho}, c \bar{c}}(p_1)
D_{\sigma \bar{\sigma}, d \bar{d}}(p_2)\\ && \times D_{\mu \bar{\mu}, a \bar{a}}(p_3) D_{\nu \bar{\nu}, b \bar{b}}(p_4)\nonumber,
\end {eqnarray}
where the momenta are subject to the constraints (\ref{D8}) and (\ref{D10}) and overall momentum 
conservation has been used ($p_4$ is a function of $p_1$, $p_2$, $p_3$.). 

For diagram B we have one four-vertex, 
two three-vertices and five propagators. Therefore, the correction to the pressure is
\begin{eqnarray}
\label{D16}
\Delta P_B &=& \frac{1}{2}\int \frac{d^4p_1d^4p_2d^4p_3}{(2\pi)^4(2\pi)^4)(2\pi)^4} \Gamma_{[4]hijk}^{\alpha \beta \gamma \lambda}\Gamma_{[3]abc}^{\mu \nu \rho}
\Gamma_{[3] \bar{a} \bar{b} \bar{c}}^{\bar{\mu} \bar{\nu} \bar{\rho}} D_{\mu \alpha, ah}(p_1)D_{\bar{\mu} \beta, \bar{a} i}(p_2) \\ && \times D_{\gamma \rho, jc}(p_3) D_{\lambda \bar{\rho}, k \bar{c}}(p_4)D_{\mu \bar{\nu}, b \bar{b}}(p_5)\nonumber,
\end{eqnarray}
where the momenta are subject to same constrains as in diagram A. It should be noticed that for the massless mode, 
$p_{5}=p_{1}-p_{3}$, we do not need to consider an additional constraint $|p_{5}|^2 \leq |\phi|^2$ because this constraint is 
automatically satisfied by the t--channel condition in (\ref{D8}). 

For diagram C, the Feynman rules are similar to diagram A and the correction to the pressure is
\begin{eqnarray}
\label{D17}
\Delta P_C &=& \frac{1}{48}\int \frac{d^4p_1d^4p_2d^4p_3}{(2\pi)^4(2\pi)^4)(2\pi)^4} \Gamma_{[4]abcd}^{\mu \nu \rho \sigma}
\Gamma_{[4]\bar{a}\bar{b}\bar{c}\bar{d}}^{\bar{\mu} \bar{\nu} \bar{\rho} \bar{\sigma}} 
D_{\rho \bar{\rho}, c \bar{c}}(p_1)D_{\sigma \bar{\sigma}, d \bar{d}}(p_2) \nonumber\\ 
&& \times D_{\mu \bar{\mu}, a \bar{a}}(p_3) D_{\nu \bar{\nu}, b \bar{b}}(p_4)\,,
\end{eqnarray}
where the momenta are subject to the constraints (\ref{D8}), (\ref{D11}), and 
(\ref{D12}).

Now, insert the Feynman rules (\ref{D1}) - (\ref{D3}) in the above integrals (\ref{D15}) - (\ref{D17}), 
consider the dimensionless variables $x_i \equiv \frac{|\vec{p_i}|}{|\phi|}$ $(i = 1, 2, 3)$ for rescaling the 
radial loop momenta, and consider the following steps:
$\\$
$\bullet$ Carry out Lorenz and color contractions.
$\\$
$\bullet$ Go over to 3D spherical coordinates. 
$\\$
$\bullet$ Trivially integrate the temporal components of four-momenta by appealing to the delta functions of the propagators.
$\\$
$\bullet$ Apply the triangle inequality to obtain upper bounds for the modulus of each correction.
$\\$
For the corrections (\ref{D15}) and (\ref{D16}) we obtain the following expressions
\begin {eqnarray}
\label{D18}
|\Delta P_{A(B)}| &\leq& (24)\,\frac{e^4 \Lambda^4\lambda^{-2}}{3\times 2^7 \times (2\pi)^6} \sum_{l,m,n=1}^2 \!\!\int dx_1 \!\!\int dx_2 \!\!\int dx_3 
\!\!\int dz_{12} \!\!\int dz_{13} \!\!\int_{z_{23,l}}^{z_{23,u}} \!\!\!\!\!\!dz_{23} \nonumber \\
&&\times \frac{1}{\sqrt{(1-z_{12}^2)(1-z_{13}^2)-(z_{23}-z_{12}z_{13})^2}}\nonumber \\
&&\times \frac{x_1^2x_2^2x_3^2}{\sqrt{x_1^2+4e^2}\sqrt{x_2^2+4e^2}\sqrt{x_3^2+4e^2}}\nonumber \\
&&\times \delta(4e^2+((-1)^{l+m}\sqrt{x_1^2+4e^2}\sqrt{x_2^2+4e^2} - x_1x_2z_{12}-\nonumber \\
&& ((-1)^{l+n}\sqrt{x_1^2+4e^2}\sqrt{x_3^2+4e^2} - x_1x_3z_{13})-\nonumber \\
&& ((-1)^{m+n}\sqrt{x_2^2+4e^2}\sqrt{x_3^2+4e^2} - x_2x_3z_{23})\nonumber \\
&&\times |P_{A(B)}(x,z,l,m,n)| n_B(2\pi\lambda^{-3/2} \sqrt{x_1^2 + 4e^2}\nonumber\times \\
&& n_B(2\pi\lambda^{-3/2} \sqrt{x_2^2 + 4e^2} n_B (2\pi\lambda^{-3/2} \sqrt{x_3^2 + 4e^2}\nonumber\times \\
&& n_B(2\pi\lambda^{-3/2} |(-1)^l \sqrt{x_1^2 + 4e^2} + (-1)^m \sqrt{x_2^2 + 4e^2}\nonumber + \\
&&(-1)^n \sqrt{x_3^2 + 4e^2}|\,, 
\end {eqnarray}
where $z_{12} \equiv \cos \angle (\vec{x_1}, \vec{x_2}), z_{13} \equiv \cos \angle (\vec{x_1}, \vec{x_3}), z_{23} \equiv 
\cos \angle (\vec{x_2}, \vec{x_3})$. The extremely lengthy polynomials $P_A$ and $P_B$ arise from Lorenz 
and color contractions and are regular at $x_1 = x_2 = x_3 = 0$. We also define \cite{Hofmann2006}:
\begin{equation}
\label{D19}
z_{23,u}\equiv\cos\left|\arccos z_{12}-\arccos z_{13}\right|\,,\ \ \ z_{23,l}\equiv\cos\left|\arccos z_{12}+\arccos z_{13}\right|\,.\\[2ex]
\end{equation}

Recalling (\ref{D10}) and the fact that only the minus sign is relevant $(e > \frac{1}{2\sqrt{2}})$ \cite{SchwarzHofmannGiacosa2007}, which makes the 
expression within the absolute-value signs strictly negative, we have
\begin{eqnarray}
\label{D20}
z_{12} & \leq & \frac{1}{x_1x_2}(4e^2 - \sqrt{x_1^2 + 4e^2} \sqrt{x_2^2 + 4e^2} + \frac{1}{2}) \equiv g(x_1, x_2) \\
\label{DD20}
z_{13} & \geq & \frac{1}{x_1x_3}(-4e^2 + \sqrt{x_1^2 + 4e^2} \sqrt{x_3^2 + 4e^2} - \frac{1}{2}) \equiv g(x_1, x_3)  \\
\label{DDD20}
z_{23} & \geq & \frac{1}{x_2x_3}(-4e^2 + \sqrt{x_2^2 + 4e^2} \sqrt{x_3^2 + 4e^2} - \frac{1}{2}) \equiv g(x_2, x_3)\,.
\end {eqnarray}
Repeating the same steps as before, for the vacuum-vacuum contribution (\ref{D17}) of diagram C we obtain
\begin {eqnarray}
\label{D21}
|\Delta P_C| & \leq & \frac{e^4\Lambda^4\lambda^{-2}}{3\times 2^5 \times (2\pi)^8} \sum_{l,m}^2 \int dy_1 \!\!\int dx_1 \!\!\int dx_2 
\!\!\int dx_3 \!\!\int dz_{12} \!\!\int dz_{13} \!\!\int_{z_{23,l}}^{z_{23,u}} \!\!\!\!\!\!dz_{23}\nonumber \\
& & \times \frac{x_1^2x_2^2x_3^2}{\sqrt{(1-z_{12}^2)(1-z_{13}^2)-(z_{23}-z_{12}z_{13})}^2}\nonumber \\
& & \times|P_{C}({\bf x},{\bf z}, y_1,l,m)| n_B(2\pi\lambda^{-3/2} \sqrt{x_3^2 + 4e^2})\nonumber \\
& & \times\frac{n_B(2\pi\lambda^{-3/2}|(-1)^l \sqrt{x_3^2 + 4e^2} + (-1)^m f_2({\bf x},{\bf z})|)}{f_2({\bf x},{\bf z}, y_1)\sqrt{x_3^2 + 4e^2}}\,,
\end {eqnarray}
where
\begin{equation}
\label{D22}
f_2({\bf x},{\bf z}) \equiv \sqrt{x_1^2 + x_2^2 +x_3^2+2x_1x_2z_{12}-2x_1x_3z_{13}-2x_2x_3z_{23}},\;\;\;\;\;\;\ y_1 \equiv \frac{p_1^{o}}{|\phi|}\,,
\end {equation}
with $z_{12}, z_{13}, z_{23}, z_{23,u}$, and $z_{23,l}$ defined in (\ref{D20}),(\ref{DD20}),(\ref{DDD20}) and (\ref{D19}), respectively. 
The polynomial $P_C$ emerges from Lorenz and color 
contractions and is regular at $x_1 = x_2 = x_3 = 0$.
According to relation (\ref{D8}) - (\ref{D11}) the rescaled constraints read:
\begin{eqnarray}
\label{D23}
|y_1^2 + y_2^2-x_1^2-x_2^2 + y_1 y_2-2x_1x_2z_{12}| &\leq & 1, \\
\label{DD23}
|y_2^2-x_2^2 + 4e^2-(-1)^l \, 2y_2 \sqrt{x_3^2+4e^2}+2x_2x_3z_{23}| &\leq & 1, \\
\label{DDD23}
|y_1^2-x_1^2 + 4e^2-(-1)^l \ 2y_1 \sqrt{x_3^2+4e^2}+2x_1x_3z_{13}| & \leq & 1, \\
\label{DDDD23}
|y_1^2 - x_1^2| \leq 1,\qquad |y_2^2 - x_2^2| &\leq& 1.
\end {eqnarray}
Moreover, $y_2$ is replaced by
\begin{equation}
\label{D24}
y_2 = -y_1 + 2(-1)^l \sqrt{x_3^2 + 4e^2} + (-1)^m f_2({\bf x}, {\bf z}) ,
\end{equation}
through delta function integration, and $(l, m = 1, 3)$.

\subsection{Monte-Carlo integration of irreducible 3-loop diagrams}

\subsubsection{Monte-Carlo integration in general}

In the following we briefly discuss how the integral of a function $f$  over a region $G \subset \mathbb{R}^n $ can 
be calculated using statistical methods. Du to the complexity of the region (algebraic varieties) $G$ that support\footnote{The region $G$ is determined by a 
set of inequalities.} 
a three-loop integration, statistical methods are required in a numerical computations of the 
values of bubble diagrams A, B, and C in Fig.\,\ref{Fig-3}. Its characteristic 
function is denoted by $\chi_{G}$. The integral then reads
\begin{equation}
\label{D25}
\int_{G} f(x) dx =\int_{\mathbb{R}^n} \chi_{G}(x)  f(x) dx.
\end {equation}
If $G$ is compact then it can be included in a box $B=[x_1,X_1] \times [x_2,X_2] \times ... \times [x_n,X_n]$. 
Therefore the integral in Eq. (\ref{D25}) can be written as
\begin {equation}
\label{D26}
\int_{\mathbb{R}^n} \chi_{G}(x)  f(x) dx =\int_B \chi_{G}(x)  f(x) dx.
\end {equation}
If $B$ has volume $V$, $\frac{1}{V} \chi_B$ can be considered as the probability density function of a 
random variable $X$, which is equally distributed thoughout the box $B$. The integral then is exactly the expectation value 
$E(V \chi_{G}(X)  f(X) )$
\begin {equation}
\label{D27}
\int_B \chi_{G}(x)  f(x) dx= V \int_B \chi_{G}(x)  f(x) \frac{1}{V} \chi_B dx = V E(\chi_{G}(X) f(X)).
\end {equation}
The Monte-Carlo method to calculate this integral consists of a statistical estimate of the expectation value. It 
is well known that the mean value $\bar{X}$  of a sample is an unbiased estimator for the expected value. If one draws 
a random sample $(x_1,x_2...,x_n)$ of points from the box, the estimate becomes:
\begin {equation}
\label{D28}
\int_{G} f(x) dx \approx V \frac{1}{n}\sum_{i=1}^n \chi_{G}(x_i)  f(x_i).
\end {equation}
The Monte-Carlo-Method is particularly useful to determine integrals over high-dimensional integration regions, 
where deterministic methods would be too time consuming. Unfortunately, the precision of this statistical estimate increases 
only with the root of the sample size. For example, this implies that the Monte-Carlo method is a relatively 
fast way to achieve a result with 
a relative precision of about 1 percent, but to achieve one more decimal place in the result the sample size must 
be increased by a factor of 100.

\subsubsection{Monte-Carlo integration for diagrams A and B}

\begin{figure}
\begin{center}
\includegraphics[scale=0.9]{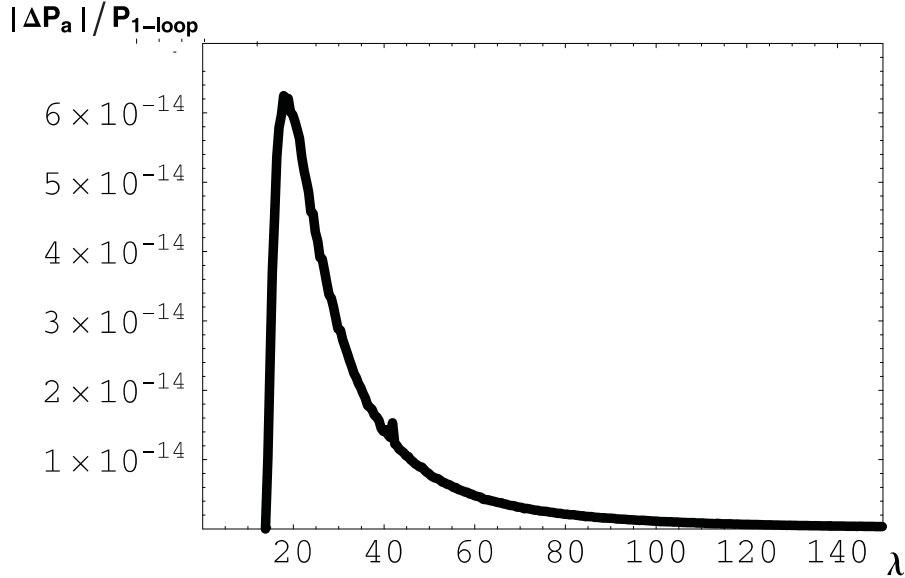}
\caption{Upper estimate for the modulus $\frac{1}{24}|\Delta P_A|/P_{\mbox{1-loop}}$ as a function of $\lambda$ for diagram A in Fig.\,\ref{Fig-3}.}
\label{Fig-4}
\end{center}
\end{figure}

\begin{figure}
\begin{center}
\includegraphics[scale=0.9]{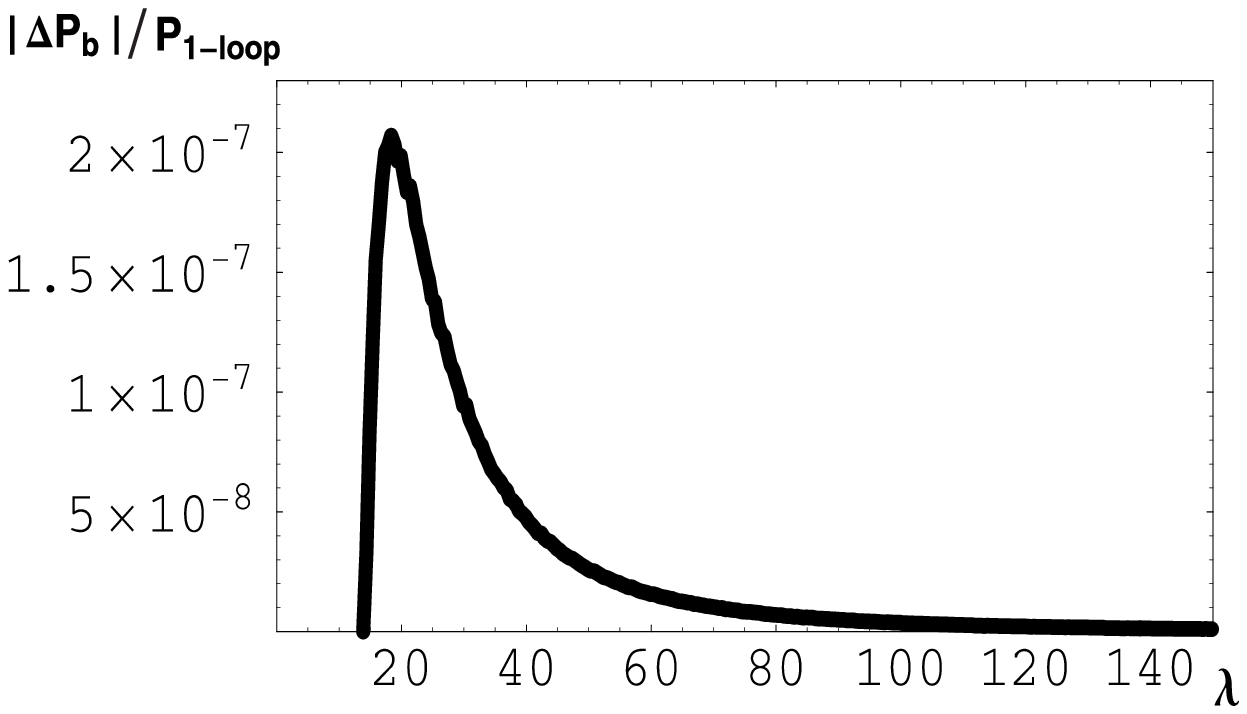}
\caption{Total upper estimate for the modulus $\frac{1}{24}|\Delta P_B|/P_{\mbox{1-loop}}$ as a function of $\lambda$ for diagram B in Fig.\,\ref{Fig-3}.}
\label{Fig-5}
\end{center}
\end{figure}

\begin{figure}
\begin{center}
\includegraphics[scale=0.9]{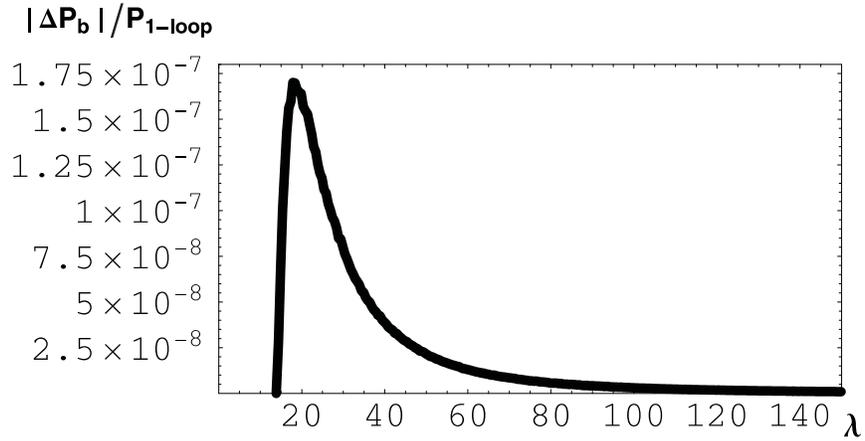}
\caption{Upper estimate for the modulus of $\frac{1}{24}|\Delta P_B|/P_{\mbox{1-loop}}$ due to the Coulomb part of the TLM propagator as a function 
of $\lambda$ for diagram B.}
\label{Fig-6}
\end{center}
\end{figure}

\begin{figure}
\begin{center}
\includegraphics[scale=0.9]{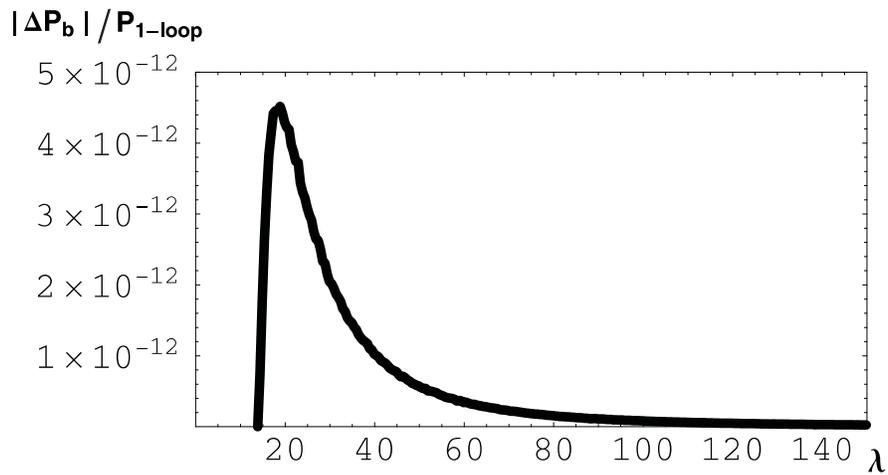}
\caption{Upper estimate for the modulus of $\frac{1}{24}|\Delta P_B|/P_{\mbox{1-loop}}$ due to TLM quantum fluctuations as a 
function of $\lambda$ for diagram B.}
\label{Fig-7}
\end{center}
\end{figure}

The calculation of $\Delta P_A$ and $\Delta P_B$ includes an integration over a 
six-dimensional region $G$ with variables of integration
\footnote{Notice that $x_1$ is integrated analytically through the delta function in $\Delta P_A$ and $\Delta P_B$.}  
$x_1,x_2,x_3,z_{12},z_{13},z_{23}$. As $z_{ij}\equiv\cos\angle(\vec{x}_i,\vec{x}_j)$, their range is restricted to the interval $[-1,1]$. 
The exact shape of $G$ is then determined by the constraints (\ref{D20}) - (\ref{DDD20}) and an additional
 angular condition (\ref{D19}) on $z_{23}$. This angular condition allows for a further reduction of the range of $z_{23}$ to the interval $[z_{23,l},z_{23,u}]$.
Also, the constraints (\ref{D20}) - (\ref{DDD20}) and condition (\ref{D20}) restrict 
the $x_i$ to the interval $[0,3]$ \cite{Hofmann2006}. Therefore a box $B$ which contains $G$ is 
identified. The constraints define the characteristic function $\chi_{G}$ whereas $z_{23,l}$ and $z_{23,u}$ are used directly as integration limits.
According to this, the ratios of the moduli $\Delta P_A$ and $\Delta P_B$ to the one-loop (ground-state subtracted) approximation 
are computed as a function of the dimensionless temperature $\lambda_c = 13.87\le\lambda\equiv\frac{2\pi T}{\Lambda}\le 140$, and 
the results are represented in Fig.\,\ref{Fig-4}--\ref{Fig-7}. For the effective gauge coupling the solution of the one-loop evolution 
equation (\ref{evalambdasu2}) is used. Figs.\,\ref{Fig-4} - \ref{Fig-7} show that at the critical 
temperature $\lambda_c = 13.87$ there is no loop correction (decoupling of TLH modes), and $\Delta P_A$ and $\Delta P_B$ approach zero for large temperatures. 
For the three-loop diagrams A and B, 
the maxima of the ratio of the estimates of their moduli to the one-loop correction are peaked between $\lambda = 17.5$ and $\lambda = 20$. They are 
$6 \cdot 10^{-14}$ and $2 \cdot 10^{-7}$, respectively. This shows the dominance of diagram B. (The next section suggests that diagram C is vanishing). 
Comparing Fig.\,\ref{Fig-5} with Fig.\,\ref{Fig-6} and Fig.\,\ref{Fig-7}, 
shows the dominance of Coulomb fluctuations $(10^{-7})$ over 
quantum fluctuation $(10^{-12})$ for in the TLM propagation contributing to diagram B.
\vspace{.5cm}
\vspace{.5cm}
\vspace{.5cm}
\vspace{.5cm}
\vspace{.5cm}
\vspace{.5cm}

\subsubsection{Monte-Carlo integration of diagram C}

The region of support for the radial loop integration the irreducible three-loop diagram C 
cannot be determined in the same way as for diagrams A and B. 
This has mainly to do with the occurrence of additional off-shell variables. 
In order to estimate the region of loop integration $G$ 
for diagram C, we consider a box with volume 
$10 \times 10 \times 10 \times 20 \times 2 \times 2 \times 2$ 
for the loop and angular variables $x_1,x_2,x_3,y_1,z_{12},z_{13},z_{23}$, respectively.  
Through the condition $|y_1^2-x_1^2| \leq 1$, and the definition 
of $z_{23,(u,l)}$ in Eq.\,(\ref{D19}) a subregion of this volume is determined, 
which roughly represents 2 \% of the volume of the box. 
In this subset $150,\!000,\!000$ points are randomly picked. Subsequently, 
it is checked whether the conditions (\ref{D23}) - (\ref{D24}). With $600,\!000,\!000$ tests we have not found a single 
point to satisfy all these conditions. This yields an estimate of a fraction of about 
$2\times 10^{-5}$ of the box volume for the region of integration. The typical distance between the sampled point 
is $0.2\sim\sqrt[7]{2 \times 10^{-5}}$. Thus it is highly probable that the region of integration is empty. 
A similar analysis applies to the vacuum-thermal combination of TLM propagators, 
and the constraints imply that the integration region is empty provided 
the region of integration for the vacuum-vacuum combination is empty. Again, we have gathered 
compelling evidence for the latter to be true.

\subsection{Comparison to 2-loops results, hierarchy}

In order to check the reliability of our numerical method in case of three-loop 
integrations, we compute the pressure of the two-loop diagram B in Fig.\,\ref{Fig-2} 
with the Monte-Carlo method and compare the result with results obtained by deterministic means 
\cite{HerbstHofmannRohrer2004,SchwarzHofmannGiacosa2007}.

\begin{figure}[!ht]
\begin{center}
\includegraphics[scale=0.7]{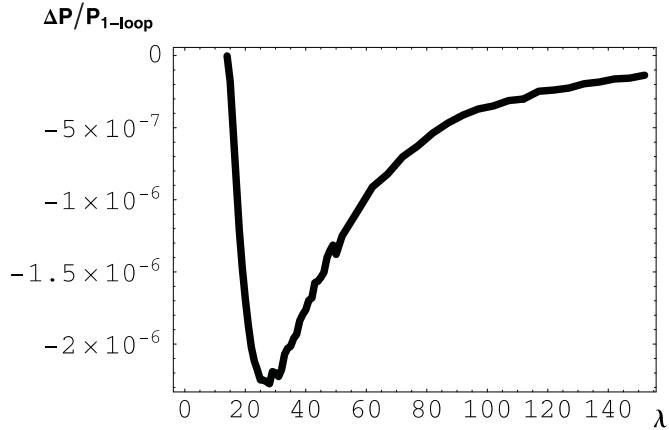}
\caption{The ratio of $\Delta P_B$ (2-loop) and $P_{1-loop}$ plotted for $13.87 \leq \lambda \leq 140$.}
\label{5_f6}
\end{center}
\end{figure}

For the two-loop diagram B in Fig.\,\ref{Fig-2}, we have two-propagators 
and one four-vertex at which the momenta 
are subject to the s--channel constraint constraint in (\ref{D8}). Thus the two-loop correction 
to the free quasiparticle pressure arising from diagram B takes the form
\begin{eqnarray}
\Delta P_{B} & = & \frac{1}{i}\int \frac{d^4p \, d^4k}{(2\pi)^8}\Gamma_{[4]abcd}^{\mu\nu\rho\delta} D_{\mu\nu,ab}(p)D_{\rho\delta,cd}(k)\,,
\end{eqnarray}
where
\begin{equation}
\label{3_g10}
|(p+k)^2| \leq {|\phi|^2}\,.
\end{equation}
As before, we can rescale the momenta as $x\equiv\frac{|\vec{p}|}{|\phi|}$ and $y\equiv\frac{|\vec{k}|}{|\phi|}$ and obtain
\begin{eqnarray}
\label{3_g11}
\Delta P_{B} & = & -\frac{e^2\Lambda^4\lambda^{-2}}{2 (2\pi)^4} \int dx \int dy \int dz_{xy} \frac{x^2y^2}{\sqrt{x^2+4e^2}\sqrt{y^2+4e^2}}\nonumber \\
&& \times P_{B}(x,y,z_{xy})n_B (2\pi\lambda^{-3/2} \sqrt{x^2+4e^2})n_B(2\pi\lambda^{-3/2}\sqrt{y^2+4e^2})\,,
\end{eqnarray}
where
\begin{equation}
\label{3_g12}
z_{xy} \leq \frac{1}{xy}(4e^2-\sqrt{x^2+4e^2}\sqrt{y^2+4e^2} + \frac{1}{2}) \equiv g(x,y)\,.
\end{equation}
Notice that 
\begin{equation}
\label{3_g14}
\lim_{x,y \to \infty} g(x,y) = -1\,.
\end{equation}
In (\ref{3_g10}) only the minus sign is relevant since $e>\frac{1}{2\sqrt{2}}$, and 
therefore the expression within the absolute-value signs is strictly negative. This leads to (\ref{3_g12}).
Apart from a small compact region, where $g(x,y) \geq 1$ and which includes the origin $x = y = 0$ in the $(x \geq 0, y \geq 0)$-quadrant, 
the admissible region of $x,y$-integration $(-1 \leq g(x,y) \leq +1)$ is an infinite strip bounded by two functions:
\begin{equation}
y^{u} (x)\equiv\frac{x+8e^2+\sqrt{1+16e^2}\sqrt{x^2+4e^2}}{8e^2}\,,
\end{equation}
\begin{equation}
y^{l} (x)\equiv\frac{x+8e^2-\sqrt{1+16e^2}\sqrt{x^2+4e^2}}{8e^2}\,.
\end{equation}
We now use the Monte-Carlo Method to calculate diagram B. 
In order to bound the region we restrict the arguments of the Bose-factors to the following interval:
\vspace{.3cm}
\begin {equation}
\label{5_g5}
0\leq 2\pi \lambda^{-3/2} \sqrt{x^2+4e^2}\leq 10
\end {equation}
and
\begin {equation}
\label{5_g6}
0\leq 2\pi \lambda^{-3/2} \sqrt{y^2+4e^2}\leq 10\,.
\end {equation}
These restrictions ensure that no Boltzmann tails are included in the region of radial loop integration. 
The integrand of (\ref{3_g11}) is plotted in Fig.\,\ref{5_f5} below. 
It is clear from Fig.\,\ref{5_f5} that the support for the $x$ and $y$ integrations is an infinite strip in the $x$-$y$ plane.
From $x = y = 150$ on suppressions are severe justifying the limits taken in (\ref{5_g5}) and (\ref{5_g6}). 
For instance, by taking $x = y = 150, e=\sqrt{8}\pi$ and $\lambda = 30$, we obtain $2\pi\lambda^{-3/2}\sqrt{x^2 + 4e^2} \sim 5.77$. 
\begin{figure}[h]
\begin{center}
\includegraphics[scale=0.7]{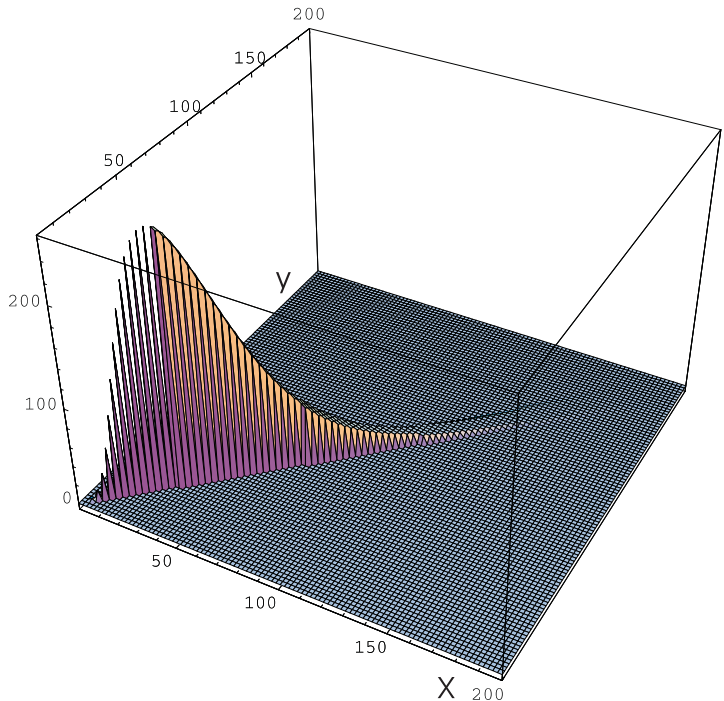}
\caption{The integrand in Eq.\,(\ref{3_g11}) plotted as a function of $x$ and $y$ for fixed $z_{xy} \equiv \cos\angle(\vec{x},\vec{y})$. The horizontal plane 
represents the $x$-$y$ plane where the integrand (mountain-shaped object) stands on an infinite strip.}
\label{5_f5}
\end{center}
\end{figure}
The ratio of two-loop to one-loop, as computed by the Monte-Carlo method, is 
represented in Fig.\,\ref{5_f6}. There are no significant deviations from the deterministic 
results computed in \cite{HerbstHofmannRohrer2004,SchwarzHofmannGiacosa2007}. Both estimates 
are of order $10^{-6}$ at temperatures $\lambda = 20 \pm 5$. 
\begin{figure}
\begin{center}
\includegraphics[scale=0.7]{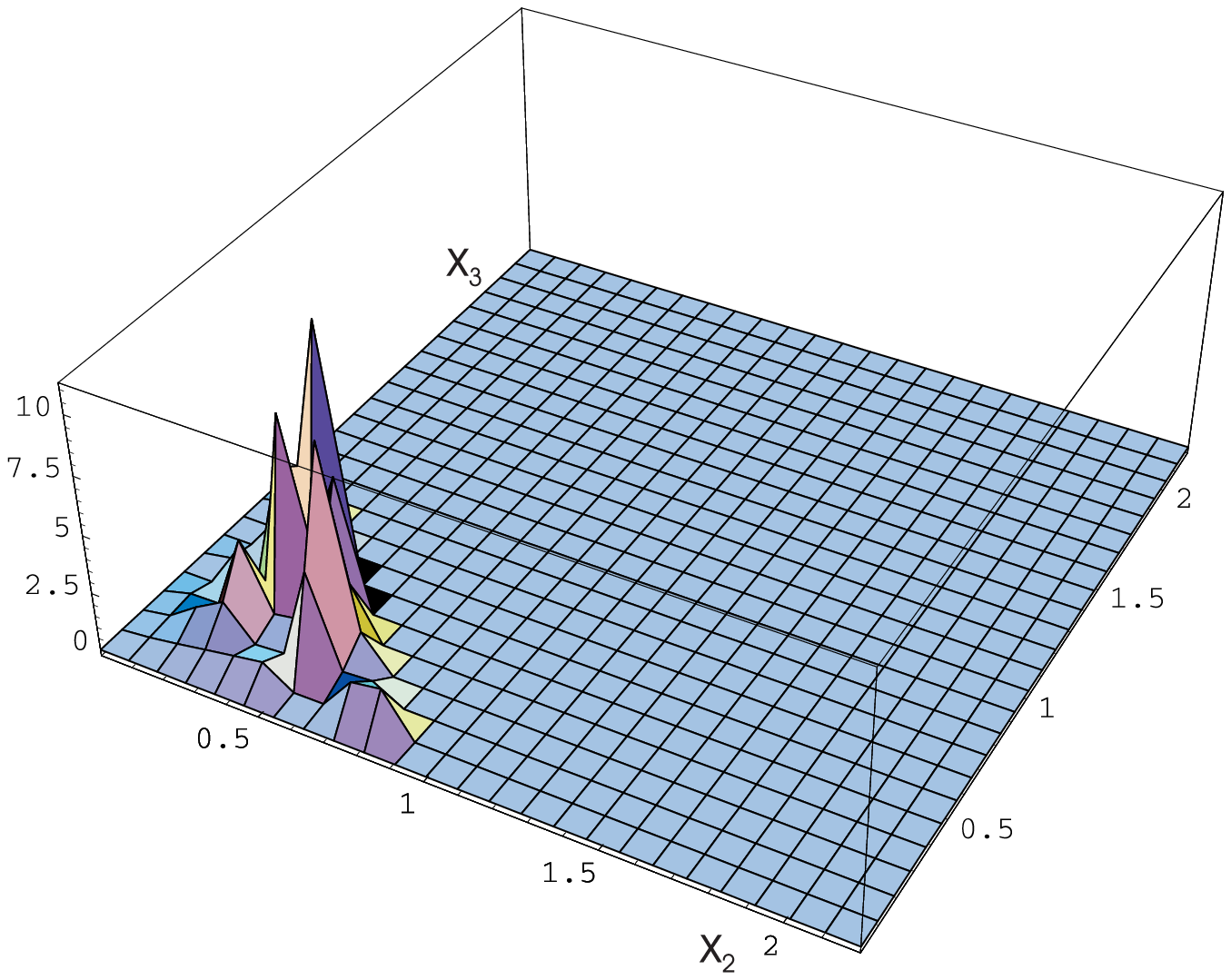}
\caption{The result of $x_1$-integration and summation over a grid of cosine 
values \mbox{$z_{12} \equiv \cos \angle (\vec{x}_1, \vec{x}_2)$}, \mbox{$z_{13} \equiv \cos \angle 
(\vec{x}_1, \vec{x}_3)$}, and \mbox{$z_{23} \equiv \cos \angle (\vec{x}_2, \vec{x}_3)$} of the integrand in 
Eq.\,(\ref{D18}) plotted as a function of $x_2$ and $x_3$, see text.} 
\label{5_f7}
\end{center}
\end{figure}
Let us now compare the integrand of (\ref{3_g11}), illustrated in Fig.\,\ref{5_f5} for the 
two-loop diagram B, with the integrand of (\ref{D18}), illustrated in Fig.\,\ref{5_f7} for the three-loop diagram B. 

To obtain an idea about the shape of the region of radial loop integration for the 
three-loop diagram B we adopt the following procedure. We use a grid of cosine values for $z_{12}$, $z_{13}$, and $z_{23}$ of 0.15 separation. This 
amounts to approximatively 2300 points. For each point 
we compute the modulus of the integrand as a function of $x_2$ and $x_3$; the $x_1$-integration can be performed trivially by appealing to the 
$\delta$-function in the integrand. 
Subsequently, a sum over all grid points 
is performed, and Fig.\,\ref{5_f7} depicts the result of this summation and the $x_1$ integration. The $x_2$-$x_3$ region 
of integration is the area where this result does not vanish. Fig.\,\ref{5_f7} supports the 
claim in \cite{Hofmann2006} that the region of 
radial loop integration in the three-loop case is compact. The fact that the upper limits of integration 
for $x_2$ and $x_3$ have been set to 3 shows that this region is fully covered in the 
Monte-Carlo process.

Recall that Fig.\,\ref{5_f5} shows the integrand for the two-loop diagram for a particular 
choice of the cosine value $z_{xy}$ close to $-1$. The region of radial loop integration is 
obviously not compact. However, the integrand is Boltzmann suppressed for large $x$ and $y$. 

The remarkable contrast between the two-loop and irreducible, planar 
three-loop diagrams can be seen from Fig.\,\ref{5_f5} and Fig.\,\ref{5_f7}: in agreement 
with our counting of constraints in Sec.\,3.2, the region of radial loop integration is 
compact in the latter while it is non-compact in the former case.
The numerical results \cite{SchwarzHofmannGiacosa2007,KavianiHofmann2007}, reviewed in Fig.\,\ref{Fig-4} - \ref{Fig-7} and Fig.\,\ref{5_f6}, 
show that one-loop, two-loop, and irreducible three-loop corrections are hierarchically ordered with a pronounced 
suppression at increasing loop order. In particular, 
comparing the modulus of the dominant irreducible three-loop correction with 
the smallest two-loop correction, we infer that they are comparable at their maxima in 
$\lambda$. On the other hand, comparing it with the dominant two-loop 
correction, shows a minimal suppression by a factor of $\sim 10^{-3}$. The dominant irreducible 
three-loop contribution is significantly dominated by Coulomb fluctuations, and the other nonvanishing 
irreducible three-loop contribution is minimally suppressed by a factor $10^{-11}$ compared to
the dominant two-loop contribution. Also, recall that one three-loop diagram appears to vanish exactly. 
These results provide numerical evidence for the conjecture \cite{Hofmann2006} on the termination of 
expansion of thermodynamical quantities or polarization tensors into irreducible loops 
(modulo resummation of reducible diagrams) to be true \footnote{As discussed in \cite{Hofmann2006}, constraints on the loop-integration variables in 
nonplanar, irreducible bubble diagrams are more severe than in the here-discussed planar case.}.

\section{Summary}

In this work we have addressed the question how the universality of the quantum of action $h$ may
find an explanation in terms of the Euclidean action of nonpropagating, selfdual gauge-field 
configurations of an SU(2) Yang-Mills theory and, after interpreting these configurations as the inducers of 
local vertices between plane waves, how the radiative correction 
of the full theory become finite. Recall that in perturbation theory, where the topologically nontrivial 
sector of field configurations is ignored, ultraviolet divergences are 
removed by a conditioned subtraction. We have pointed out that this pragmatism of renormalization theory, 
which was proved a long time ago to be physically consistent \cite{Veltman'Hooft,LeeZinnJustin}, is justified by solely a small window 
in scale parameter, associated with the just-not-resolved, relevant topological sector of the theory 
(calorons/anticalorons of topological charge modulus unity), 
inducing local vertices between plane waves. We also have discussed 
how the coupling of photons to electric charges is related to the dual of the effective gauge coupling in the 
deconfining phase. Finally, to illustrate the workings of effective loops 
we have made explicit certain technicalities of the effective pressure expansion in deconfining SU(2) Yang-Mills thermodynamics. Numerically, the 
hierarchical suppression of 
increasing loop orders supports the conjecture of a termination at a finite order of the 
expansion into irreducible diagrams \cite{Hofmann2006}.

\end{document}